\documentclass[aps,prl,reprint,longbibliography]{revtex4-1}

\usepackage{times}
\usepackage{amsmath}
\usepackage{graphicx}
\usepackage{verbatim}
\usepackage{color}
\usepackage[colorlinks=false]{hyperref}
\begin{document}

\title{Refining Glass Structure in Two Dimensions}
\author{Mahdi Sadjadi}
\email[]{mahdisadjadi@asu.edu}
\affiliation{Department of Physics,
Arizona State University, Tempe, AZ 85287-1604}

\author{Bishal Bhattarai}
\email[]{bb248213@ohio.edu}
\affiliation{Department of Physics and Astronomy, Ohio University, Athens,
OH 45701, United States}

\author{D.A. Drabold}
\email{drabold@ohio.edu}
\affiliation{Department of Physics and Astronomy, Ohio University, Athens,
OH 45701, United States}

\author{M.F. Thorpe}
\email[]{mft@asu.edu}
\affiliation{Department of Physics,
Arizona State University, Tempe, AZ 85287-1604}
\affiliation{Rudolf Peierls Centre for Theoretical Physics,
University of Oxford, 1 Keble Road, Oxford OX1 3NP, U.K}

\author{Mark Wilson}
\email[]{mark.wilson@chem.ox.ac.uk}
\affiliation{Department of Chemistry, Physical and Theoretical Chemistry Laboratory,
University Of Oxford, South Parks Road,Oxford OX1 3QZ, U.K}

\begin{abstract}
Recently determined atomistic scale structures of near-two dimensional
bilayers of vitreous silica (using scanning probe and electron microscopy)
allow us to refine the experimentally determined coordinates to incorporate the
known local chemistry more precisely.  Further refinement is achieved by using classical
potentials of varying complexity;  one using harmonic potentials and the second
employing an electrostatic description incorporating polarization effects. These
are benchmarked against density functional calculations. Our main
findings are that (a) there is a symmetry plane between the two disordered layers;
a nice example of an emergent phenomenon, (b) the layers are slightly tilted so
that the Si-O-Si angle between the two layers is not $180^{\circ}$ as originally
thought but rather $175 \pm 2 ^{\circ}$ and (c) while interior areas that are not
completely imagined can be reliably reconstructed, surface areas are more
problematical.  It is shown that small crystallites that appear are just as
expected statistically in a  continuous random network. This provides a good
example of the value that can be added to disordered structures imaged at the
atomic level by implementing computer refinement.
\end{abstract}

\maketitle

The atomic structure of covalent network glasses has been a subject of both
experimental and theoretical interest since the introduction of the Continuous Random Network
(CRN) model by Zachariasen~\cite{zachariasen1932}.
Almost all of these studies have focused on the Pair Distribution Function (PDF)
which is the Fourier transform of a diffraction pattern~\cite{wright2013}.
Experimental diffraction studies offer useful information, in particular
regarding pair-wise ordering~\cite{fischer2005}. However, simulation models can greatly aid
the interpretation of these data as the atom positions are known
unequivocally. As a result, information such as the ring statistics, which is in
many ways a natural language for discussing network
structure~\cite{marians1990,zeidler2014,sadjadi2016},
is directly accessible.
While this work has been very informative and clearly established the correctness
of the CRN model for materials like vitreous silica, it is not accurate enough
to distinguish between different models with varying ring statistics \textit{etc}.
This situation has changed recently with the direct imaging of bilayers of
silica~\cite{lichtenstein2012,huang2012} that has provided detailed
information regarding atomic positions.

\begin{figure}[b]
  {\includegraphics[scale=0.20]{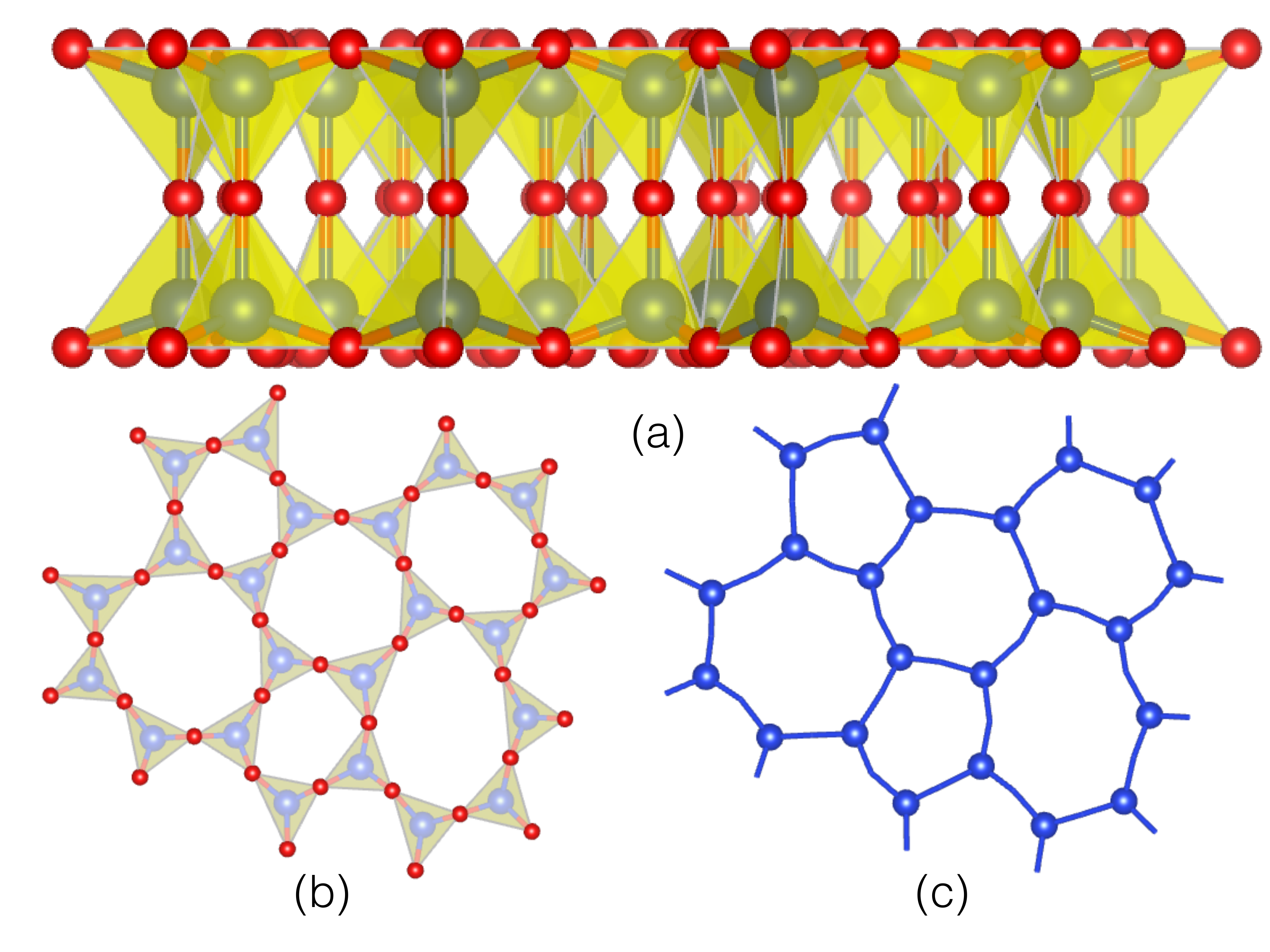}}
  \caption{ \label{fig:network} (a) A small piece of silica bilayer in which oxygen atoms (red)
  form a tetrahedral network while silicons (blue) are located at the center of
  tetrahedra. (b) The top view of the silica bilayer
  where O and Si atoms are projected into the plane, with O
  forming a network of corner-sharing triangles. (c) An alternative view where
  Si atoms form a network of edge-sharing polygons (rings), while oxygens are removed
  for clarity. This view is stressed in Fig.~\ref{fig:topview}.}
\end{figure}

\begin{figure}[t]
\centering
{\includegraphics[scale=0.25]{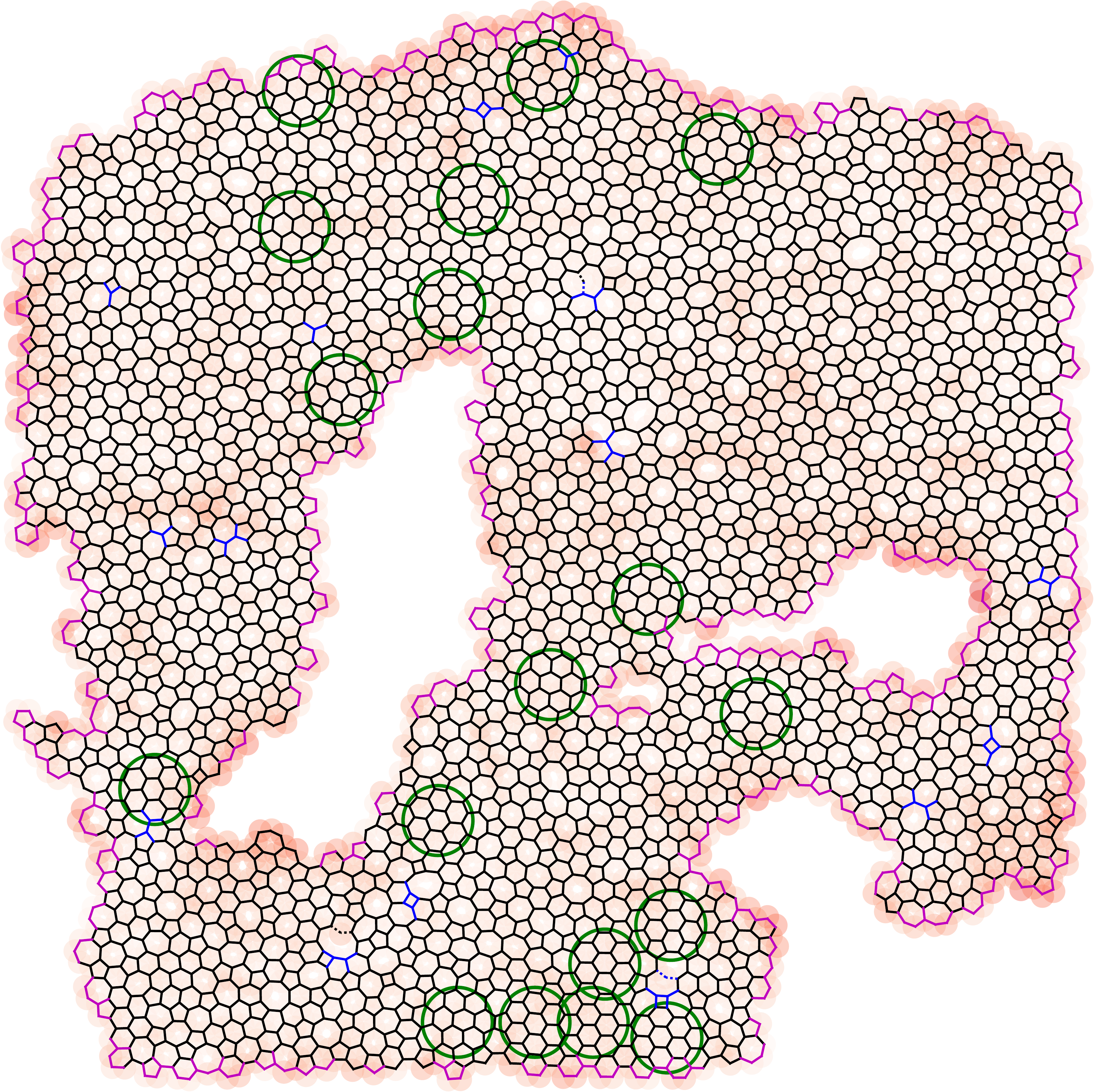}}
\caption{\label{fig:topview} The Cornell {\it h} network viewed perpendicular to the
plane containing the bilayer with the O atoms removed for clarity, and only the top layer
of Si atoms shown as {\it vertices}. The atoms associated with the blue and magenta bonds were
not directly imaged but have been added in the computer refinement. The blue and
magenta bonds highlight bonds reconstructed within the main body of the sample
and at the surface, respectively. Dashed lines highlight small sections in which
an under-coordinated Si atom was required for filling. The intensity of the red
highlights the difference between the configuration relaxed with the spring and PIM
potentials. The green circles show small {\it crystallites}.}
\end{figure}

Silica, SiO$_2$, represents an archetypal network-forming material. At ambient
pressure the crystalline and amorphous structures can
be considered as constructed from corner-sharing SiO$_4$ tetrahedral
coordination polyhedra (CP) which link to form a network.
The complex linking of the CP may result in
significant ordering on length-scales beyond the
short-range ordering imposed by the system electrostatics (effectively
controlled by the relative atom
electronegativities)~\cite{salmon2005a,salmon2005b,salmon2007b,salmon2006,salmon2007,wilson2016}.

Recently developed synthetic pathways have allowed thin films of SiO$_2$ to be
deposited on either metallic~\cite{weissenrieder2005,loffler2010,lichtenstein2012}
or graphitic~\cite{huang2012} substrates whilst
advances in imaging techniques allow for true atomic resolution of the surface structure.
Albeit, because the bilayer is a glassy material, it is not commensurate with
any substrate, and so we do not include the substrate here.

Some of the thinnest films deposited are bilayers of corner-sharing SiO$_4$ CP in
which all of the Si and O atoms obtain
their full (four- and two- respectively) coordination numbers. Amorphous and crystalline films
have been grown with both states characterized by the presence of a mirror plane
(which houses a layer of O atoms which act as bridges between the two
monolayers~\cite{wilson2013}). Critically, the pseudo-two-dimensional nature of these
systems allows the ring structures to be directly observed and hence offers
a potentially unique insight into the origin of any ordering on long length-scales.
Silica can be considered as a network of silicon atoms in which the nearest-neighbor
Si-Si pairs are {\it dressed} with O atoms. As a result, the crystalline system can be
considered as constructed exclusively from
a net of six-membered (Si-Si-Si...) rings, whilst the amorphous systems
are constructed from a distribution of 4- to 10-membered rings (Fig.~\ref{fig:network}).
However, this new experimental
information, whilst ground-breaking, is naturally imperfect as the location of
each atom has associated with it a natural
uncertainty which translates into an uncertainty in atom-atom separations.

In this Letter, we show how value can be added by combining
the experimental image with computer refinement that builds in the known local chemistry.
Whilst no refinement of the experimental data is required in order to obtain, for example,
accurate ring statistics, refinement is required in order to address the geometrical issues associated with
the network. For example, value can be added on the effect of the presence of significant unimaged regions
as well as on the subtle variations in the structure perpendicular to the resolved
plane containing the bilayer.

In this Letter we focus on a single large sample of a bilayer of vitreous silica
imaged by the Cornell group~\cite{huang2012} which we will refer to as sample {\it{h}},
shown in Fig.~\ref{fig:topview}, to distinguish it from previous smaller experimental
and computer-generated samples~\cite{kumar2014}. The sample is
~$\sim{270}\times{270}{\textrm{\AA}}^2$ in area containing 19,330 O and 9,492
Si atoms, and is the largest such sample imaged at the atomic level of which we
are aware.

Importantly, we are using the whole experimental sample, including voids,
rather than selecting a more rectangular shaped section without voids, which
would have thrown out most of the experimental data. This is also significant as the full
configuration shows a number of interesting features. For example, there
are several regions which may be considered nanocrystalline
showing relatively large numbers of neighbouring six-membered rings
(highlighted by green circles with a diameter of 9 {\AA}).  Such regions are to be
expected staistically in a CRN and from previous studies~\cite{sadjadi2016} we find
that about 50\% of all
rings are sixfold and of these about 2\% are surrounded by 6 sixfold rings leading to
a little {\it{microcrystallite}} of 7 sixfold rings. The total number of rings in
the Cornell {\it{h}} sample is 1811, where we exclude surface rings that do not have
their full compliment of neigboring rings. Thus we expect 1811 $\times$ 0.5 $\times$ 0.02 $\approx$ 18 of
such regions which is fortuitously exactly the number of regions shown by green circles.
So this certainly cannot be taken as any evidence for microcrystallites as has been
postulated at various times since the original ideas of Lebedev and coworkers~\cite{lebedev1937}.

More obviously the configuration shows three relatively large regions which were
unable to be imaged (of approximate dimensions  160$\times$40${\textrm{\AA}}^2$,
50$\times$20${\textrm{\AA}}^2$ and 10$\times$10${\textrm{\AA}}^2$ respectively)
which resist reasonable attempts at computational filling (see below). A
potential implication is that the underlying surface (on which the bilayer
has been grown) in some way distorts the bilayer thus preventing effective
imaging or perhaps the network was never formed in these regions because of
surface roughness.

To construct the bilayer from the experimental image, O atoms (which are not imaged)
are placed midway between Si atoms (which are imaged) thus forming a network of
corner-sharing O$_3$ triangles (each of which has an Si atom at the centre).
The Si and O atoms planes are then separated, forming trigonal pyramids with Si atoms
at the apices.
A mirror image of these pyramids is joined to the original via O-atom bridges to
form the completed bilayer, resulting in an initial set of 180$^o$ Si-O-Si bond
angles centered around the O atoms in the mirror plane (Fig.~\ref{fig:network}).
An important question involves the experimental length metric to ensure the correct
calibration of the image. We calculated the mean average length of the imaged
nearest neighbour Si-Si distances as 3.097\AA{}, which is close to the expected value of
3.100\AA{} for glassy silica structures~\cite{keen1999}, confirming the overall accuracy of the experiment,
and alleviating the need for any length rescaling. To reconstruct the unimaged
regions, we use mean bond length and internal angles of rings to
find the correct local topology. The subsequent relaxation of the bilayer
will fix the geometry ensuring the proper bond length and angles.

This relaxation is carried out using model potentials of increasing complexity.
In the simplest case, the nearest-neighbour O-O bonds are
mimicked by harmonic springs with lengths set as the mean average ($2.645${\AA}).
This ensures that the system does not have any internal degrees of freedom
and is minimally rigid or isostatic~\cite{thorpe1983continuous,ellenbroek2015rigidity}.
A hardcore potential is added to prevent overlap of O atoms from different tetrahedra
as well as an RMSD (Root-Mean-Square Deviation) term which penalizes deviation from
the experimental coordinates. This RMSD term involves the sum of squares of the
refined minus the experimental atomic positions and is important as this maintains
the overall area and alleviates the need for additional boundary conditions to
maintain the sample area. Although proper boundary conditions for finite pieces
of amorphous systems can be designed \cite{theran2015}, this simple potential
can account for structural information extracted in this Letter.
Maintaining the configurational area is critical in avoiding, for example,
unphysical overlaps in nearest-neighbour tetrahedra in the absence of formal
(electrostatic) repulsions. The balance of the surface extension and the
inter-tetrahedral repulsions define an effective flexibility window of acceptable
structural solutions, of the type commonly associated with zeolites \cite{sartbaeva2007}.
As a result, samples with irregular boundary conditions are not a problem.

\begin{figure}[t!]
\centering
\includegraphics[bb= 27 32 706 583, scale=0.3]{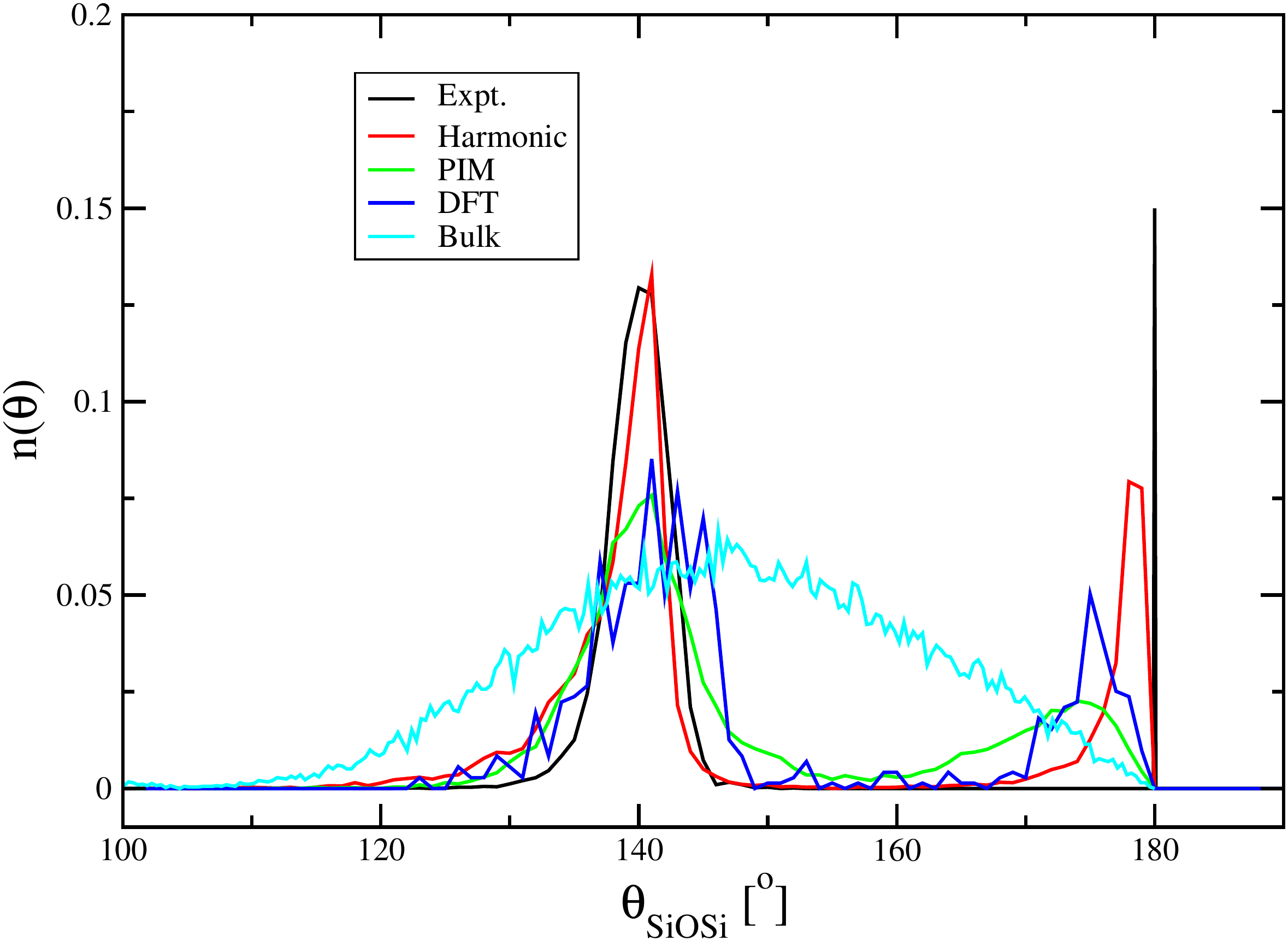}
\caption{\label{fig:tiltangle} The Si-O-Si bond angle distributions determined from the
original experimental configuration
and from the bilayers obtained using models of increasing complexity as well as for the
 bulk glass. The peak at $\theta_{\textrm{SiOSi}}\sim{145}^{\circ}$ arises
from the ``in-plane'' tetrahedral links whilst the peak at $\sim{180}^{\circ}$ arises
from the central bridging oxygen atoms between the two planes.
The unrefined experimental result for the Cornell {\it{h}} sample is shown in black where it
was \textit{assumed} that the central bridging angle was exactly $180^{\circ}$.
The DFT calculation is on a computer-generated periodic sample and acts as the best
guide for what to expect.  The other two results are for the
refined Cornell {\it{h}} sample using both the harmonic model and the polarizable-ion
model as described in the text. Both show significant tilting as expected from
the results of DFT, while maintaining the central symmetry plane.}
\end{figure}

A second classical model used is a polarizable-ion model (PIM) \cite{madden1996}, specifically
the TS potential \cite{tangney2002} which utilises pair potentials to model the Coulomb,
short-range (overlap) and dispersive interactions. The potential employs a combination of reduced
ion charges and anion dipole polarisation (as described in reference \cite{madden1996}).
The results from the harmonic potential model are used as the input with the PIM
further refining the results.

The most sophisticated method applied uses Density Functional Theory (DFT). However,
the method is too computationally-demanding to apply to the experimental Cornell
{\it{h}} configuration. Therefore, a relatively small $1200$ atom periodic
computer-generated model (with 200 Si atoms in each monolayer)
of a vitreous silica bilayer~\cite{kumar2012} was used. Density
functional calculations were undertaken with the code SIESTA~\cite{soler2002},
with single-zeta basis and the local density approximation. Relaxation
with a variable cell area resulted in very little change. Stability
of the relaxed model was also verified~\cite{bhattarai2016}.

The result of the PIM refinement of Cornell {\it h} is shown in Fig.~\ref{fig:topview}.
The blue (bulk) and magenta (surface) bonds have been computer-reconstructed,
as described earlier. The interior reconstruction was deemed to be successful,
as the differences between the spring and PIM models were minor.
These differences are shown by the red shading where the darkest red corresponds
to an atomic displacement of $\sim{0.5}$\AA{} from the original (unrefined) coordinates.
This strongly suggests that the network existed in these interior areas but was not
imaged reliably, rather than the networks growing around a pillar or avoiding surface
roughness on the substrate and never existing. At the surface, the difference between
the spring and PIM models was much greater as the reconstruction was not contained
within a small closed exterior perimeter.

In addition to in-plane information, refinement can provide valuable information
in perpendicular direction. As a benchmark of our model potentials,
we have studied the Si-O-Si angle, $\theta_{\textrm{SiOSi}}$, as this contains
important information on how the tetrahedra are linked. Figure
\ref{fig:tiltangle} shows the distributions of $\theta_{\textrm{SiOSi}}$
for three models. The experimental structure (in which
linear Si-O-Si bridges between the two monolayers are imposed) shows
a bond angle of $\theta_{\textrm{SiOSi}}\sim{140.3}^o$
(with a FWHM of $\Delta\theta\sim{5.2}^\circ$)
in the bilayer plane. All of the models generate bimodal distributions in
which the peak at $\theta_{\textrm{SiOSi}}\sim{145}^o$ may be assigned to the
Si-O-Si triplets in the bilayer plane whilst the peaks at
$\theta_{\textrm{SiOSi}}>175^o$ correspond to the triplets centred around the
bridging O atoms in the mirror plane (\textit{i.e.} perpendicular to the bilayer plane).
There is not much latitude in the in-plane values of this angle as they must be consistent
with the measured area and the known Si-O bond lengths, which leads to a single peak in the
$\theta_{\textrm{SiOSi}}\sim{145^{\circ}}$. Fig.~\ref{fig:tiltangle} shows
that the harmonic model reproduces the important high-angle
peak at $\theta\sim{178.5}^\circ$.  The lower-angle peak
is at $\theta\sim{140.9}^\circ$ ($\Delta\theta\sim{3.8}^\circ$)
and some way below the DFT result.

The figure also shows the analogous distribution obtained from the bulk glass at
ambient pressure using PIM, which is similar to distributions
observed in bulk silicates~\cite{desa1982,neuefeind1996}. The bulk distribution is
significantly broader than those generated for the bilayer with
$\theta\sim{145}^\circ$ and $\Delta\theta\sim{36}^\circ$.
The requirement for the relatively obtuse bond angles which
characterise the links between the two layers constrains the in-plane
bond angles to a relatively narrow range. For the intra-layer angles all of
the models show peaks at $\theta\sim{140}-142^\circ$ with the harmonic potential
showing a far sharper peak retaining the symmetry plane.

\begin{figure}[t!]
  \centering
  {\includegraphics[scale=0.07]{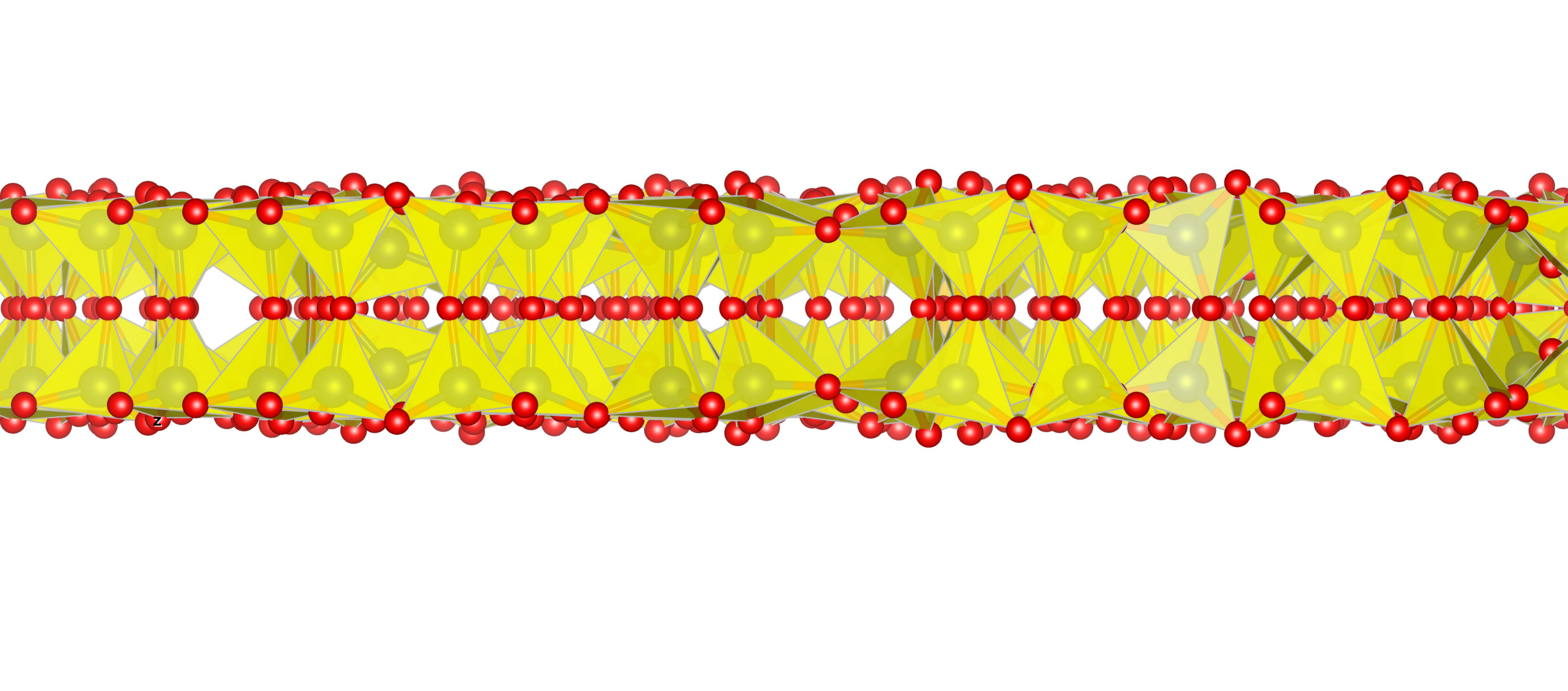}}
  \caption{\label{fig:sideview} A section of the Cornell {\it{h}} network shown along the plane
  containing the bilayer with O atoms shown in red, and with
  Si atoms at the center of the yellow tetrahedra. Note the symmetry plane of the central
  O atoms and also the tilting of the tetrahedra away from the vertical about the central plane.}
\end{figure}

However the bridging O angle is tilted and reduced to about $175.1^{\circ}$. A Si-O-Si
angle of 180$^o$ sits on a local energy maximum \cite{dawson2014} and, as a result,
tilting is inevitable. A tilt in the inter-layer bond angle is observed in all the models.
At the simplest level (harmonic potential) a relatively small deviation from linear
($\theta\sim{178.5}^\circ$) is shown. As greater detail is added to the models
these angles become more acute with both the PIM and DFT results showing peaks
at $\theta\sim{175}^o$.  Figure \ref{fig:sideview} shows the
configuration perpendicular to the plane containing the bilayer relaxed using the PIM
and clearly showing the tilted corner-sharing tetrahedra, with a peak at $\theta\sim{174.9}^\circ$.

At first sight this suggests
an incompatibility with the experimental results where only a
single layer is seen, with the second layer of Si tetrahedra being exactly
behind and underneath the first. However this can be maintained if there is a
symmetry plane involving the central O atoms, such that the upper and lower
tetrahedra tilt and pucker in the same way and there is not a second image when the
bilayer is imaged from above, as shown in Figure \ref{fig:sideview}.
This conclusion is supported by an entropy argument in which
the bilayer \textit{with} a mirror plane is able to explore configurational space
more effectively than one without~\cite{wilson2013}. There are more degrees of
freedom with the symmetry plane present, thus increasing the
entropy and lowering the free energy, and hence leading to this unexpected emergent phenomena.
Thus symmetry is induced in a system which at first sight seems a canonical example
of a system without symmetry. This argument is confirmed both by detailed atomic computer
modeling and by experiment, where no {\it{shadow}} is seen beside each atom imaged,
so that the second layer must be exactly behind the first layer.

A feature to notice from Fig.~\ref{fig:topview} is that the polygons with silicon atoms
at the corners appear regular, having areas close to that of regular polygons
as has been previously noted~\cite{kumar2014}.  This feature has been absent in computer
generated models of vitreous silica bilayer as the Si-O-Si angle of around
$145^{\circ}$ in the plane is hard to achieve in models while maintaining
the maximal convexity of Si polygons. Nature has found a way and we need to understand better
how this is achieved. Note there is no difficulty in achieving regular polygons in samples
of amorphous As~\cite{kumar2012} where there are no bridging atoms to contend with.

In this Letter we have described how computer-refinement can add value to experimental
images of disordered structures at the atomic level.
Although this is the first time this has been attempted with an amorphous structure,
 with advances in imaging, many more such systems are expected to be imaged in the near future. This
somewhat parallels the procedures employed to rationalise protein structure where the
local chemistry, via bond lengths \textit{etc} is included to produce the best possible
structure~\cite{perrakis1999}. We have shown that simple potentials are adequate here,
and as well as producing refined coordinates for the bilayer (available upon request),
we have shown that the two layers are tilted while maintaining a flat central symmetry
plane of O atoms between the upper and lower parts of the bilayer. It is remarkable
that such symmetry can exist in disordered system and this can be viewed as a nice clean
example of an emergent phenomena.

Future work will help determine how ubiquitous bilayer structures of this type may be. It is
possible, for example, that forming such structures for systems such as GeO$_2$ may be more
problematic as a significantly larger tilt ($\theta\ll{180}^\circ$) would have to
be accommodated \cite{dawson2014}.

\begin{acknowledgments}
  We should like to thank Berlin and Cornell groups for the coordinates of their
  networks and for useful discussions. This work used the Extreme Science and
  Engineering Discovery Environment (XSEDE), which is supported by National
  Science Foundation grant number ACI-1548562~\cite{towns2014}.
  Support through NSF grants~\# DMS 1564468 (MFT) and
  ~\# DMR 1506836 (DAD) is gratefully acknowledged.
\end{acknowledgments}

\bibliography{reference}

\begin{thebibliography}{34}%
\makeatletter
\providecommand \@ifxundefined [1]{%
 \@ifx{#1\undefined}
}%
\providecommand \@ifnum [1]{%
 \ifnum #1\expandafter \@firstoftwo
 \else \expandafter \@secondoftwo
 \fi
}%
\providecommand \@ifx [1]{%
 \ifx #1\expandafter \@firstoftwo
 \else \expandafter \@secondoftwo
 \fi
}%
\providecommand \natexlab [1]{#1}%
\providecommand \enquote  [1]{``#1''}%
\providecommand \bibnamefont  [1]{#1}%
\providecommand \bibfnamefont [1]{#1}%
\providecommand \citenamefont [1]{#1}%
\providecommand \href@noop [0]{\@secondoftwo}%
\providecommand \href [0]{\begingroup \@sanitize@url \@href}%
\providecommand \@href[1]{\@@startlink{#1}\@@href}%
\providecommand \@@href[1]{\endgroup#1\@@endlink}%
\providecommand \@sanitize@url [0]{\catcode `\\12\catcode `\$12\catcode
  `\&12\catcode `\#12\catcode `\^12\catcode `\_12\catcode `\%12\relax}%
\providecommand \@@startlink[1]{}%
\providecommand \@@endlink[0]{}%
\providecommand \url  [0]{\begingroup\@sanitize@url \@url }%
\providecommand \@url [1]{\endgroup\@href {#1}{\urlprefix }}%
\providecommand \urlprefix  [0]{URL }%
\providecommand \Eprint [0]{\href }%
\providecommand \doibase [0]{http://dx.doi.org/}%
\providecommand \selectlanguage [0]{\@gobble}%
\providecommand \bibinfo  [0]{\@secondoftwo}%
\providecommand \bibfield  [0]{\@secondoftwo}%
\providecommand \translation [1]{[#1]}%
\providecommand \BibitemOpen [0]{}%
\providecommand \bibitemStop [0]{}%
\providecommand \bibitemNoStop [0]{.\EOS\space}%
\providecommand \EOS [0]{\spacefactor3000\relax}%
\providecommand \BibitemShut  [1]{\csname bibitem#1\endcsname}%
\let\auto@bib@innerbib\@empty
\bibitem [{\citenamefont {Zachariasen}(1932)}]{zachariasen1932}%
  \BibitemOpen
  \bibfield  {author} {\bibinfo {author} {\bibfnamefont {W.H.}\ \bibnamefont
  {Zachariasen}},\ }\bibfield  {title} {\enquote {\bibinfo {title} {The atomic
  arrangement in glass},}\ }\href {\doibase 10.1021/ja01349a006} {\bibfield
  {journal} {\bibinfo  {journal} {Journal of the American Chemical Society}\
  }\textbf {\bibinfo {volume} {54}},\ \bibinfo {pages} {3841--3851} (\bibinfo
  {year} {1932})}\BibitemShut {NoStop}%
\bibitem [{\citenamefont {Wright}\ and\ \citenamefont
  {M.F.Thorpe}(2013)}]{wright2013}%
  \BibitemOpen
  \bibfield  {author} {\bibinfo {author} {\bibfnamefont {A.C.}\ \bibnamefont
  {Wright}}\ and\ \bibinfo {author} {\bibnamefont {M.F.Thorpe}},\ }\bibfield
  {title} {\enquote {\bibinfo {title} {Eighty years of random networks},}\
  }\href {\doibase 10.1002/pssb.201248500} {\bibfield  {journal} {\bibinfo
  {journal} {physica status solidi (b)}\ }\textbf {\bibinfo {volume} {250}},\
  \bibinfo {pages} {931--936} (\bibinfo {year} {2013})}\BibitemShut {NoStop}%
\bibitem [{\citenamefont {Fischer}\ \emph {et~al.}(2005)\citenamefont
  {Fischer}, \citenamefont {Barnes},\ and\ \citenamefont
  {Salmon}}]{fischer2005}%
  \BibitemOpen
  \bibfield  {author} {\bibinfo {author} {\bibfnamefont {Henry~E}\ \bibnamefont
  {Fischer}}, \bibinfo {author} {\bibfnamefont {Adrian~C}\ \bibnamefont
  {Barnes}}, \ and\ \bibinfo {author} {\bibfnamefont {Philip~S}\ \bibnamefont
  {Salmon}},\ }\bibfield  {title} {\enquote {\bibinfo {title} {Neutron and
  x-ray diffraction studies of liquids and glasses},}\ }\href {\doibase
  10.1088/0034-4885/69/1/R05} {\bibfield  {journal} {\bibinfo  {journal}
  {Reports on Progress in Physics}\ }\textbf {\bibinfo {volume} {69}},\
  \bibinfo {pages} {233} (\bibinfo {year} {2005})}\BibitemShut {NoStop}%
\bibitem [{\citenamefont {Marians}\ and\ \citenamefont
  {Hobbs}(1990)}]{marians1990}%
  \BibitemOpen
  \bibfield  {author} {\bibinfo {author} {\bibfnamefont {Carol~S}\ \bibnamefont
  {Marians}}\ and\ \bibinfo {author} {\bibfnamefont {Linn~W}\ \bibnamefont
  {Hobbs}},\ }\bibfield  {title} {\enquote {\bibinfo {title} {Local structure
  of silica glasses},}\ }\href {\doibase 10.1016/0022-3093(90)90299-2}
  {\bibfield  {journal} {\bibinfo  {journal} {Journal of Non-Crystalline
  Solids}\ }\textbf {\bibinfo {volume} {119}},\ \bibinfo {pages} {269--282}
  (\bibinfo {year} {1990})}\BibitemShut {NoStop}%
\bibitem [{\citenamefont {Zeidler}\ \emph {et~al.}(2014)\citenamefont
  {Zeidler}, \citenamefont {Wezka}, \citenamefont {Rowlands}, \citenamefont
  {Whittaker}, \citenamefont {Salmon}, \citenamefont {Polidori}, \citenamefont
  {Drewitt}, \citenamefont {Klotz}, \citenamefont {Fischer}, \citenamefont
  {Wilding} \emph {et~al.}}]{zeidler2014}%
  \BibitemOpen
  \bibfield  {author} {\bibinfo {author} {\bibfnamefont {Anita}\ \bibnamefont
  {Zeidler}}, \bibinfo {author} {\bibfnamefont {Kamil}\ \bibnamefont {Wezka}},
  \bibinfo {author} {\bibfnamefont {Ruth~F}\ \bibnamefont {Rowlands}}, \bibinfo
  {author} {\bibfnamefont {Dean~AJ}\ \bibnamefont {Whittaker}}, \bibinfo
  {author} {\bibfnamefont {Philip~S}\ \bibnamefont {Salmon}}, \bibinfo {author}
  {\bibfnamefont {Annalisa}\ \bibnamefont {Polidori}}, \bibinfo {author}
  {\bibfnamefont {James~WE}\ \bibnamefont {Drewitt}}, \bibinfo {author}
  {\bibfnamefont {Stefan}\ \bibnamefont {Klotz}}, \bibinfo {author}
  {\bibfnamefont {Henry~E}\ \bibnamefont {Fischer}}, \bibinfo {author}
  {\bibfnamefont {Martin~C}\ \bibnamefont {Wilding}},  \emph {et~al.},\
  }\bibfield  {title} {\enquote {\bibinfo {title} {High-pressure transformation
  of {S}i{O}$_2$ glass from a tetrahedral to an octahedral network: a joint
  approach using neutron diffraction and molecular dynamics},}\ }\href
  {\doibase 10.1103/PhysRevLett.113.135501} {\bibfield  {journal} {\bibinfo
  {journal} {Physical review letters}\ }\textbf {\bibinfo {volume} {113}},\
  \bibinfo {pages} {135501} (\bibinfo {year} {2014})}\BibitemShut {NoStop}%
\bibitem [{\citenamefont {Sadjadi}\ and\ \citenamefont
  {M.F.Thorpe}(2016)}]{sadjadi2016}%
  \BibitemOpen
  \bibfield  {author} {\bibinfo {author} {\bibfnamefont {M.}~\bibnamefont
  {Sadjadi}}\ and\ \bibinfo {author} {\bibnamefont {M.F.Thorpe}},\ }\bibfield
  {title} {\enquote {\bibinfo {title} {Ring correlations in random networks},}\
  }\href {\doibase 10.1103/PhysRevE.94.062304} {\bibfield  {journal} {\bibinfo
  {journal} {Physical Review E}\ }\textbf {\bibinfo {volume} {94}},\ \bibinfo
  {pages} {062304} (\bibinfo {year} {2016})}\BibitemShut {NoStop}%
\bibitem [{\citenamefont {L.Lichtenstein}\ \emph {et~al.}(2012)\citenamefont
  {L.Lichtenstein}, \citenamefont {Buechner}, \citenamefont {B.Yang},
  \citenamefont {S.Shaikhutdinov}, \citenamefont {M.Heyde}, \citenamefont
  {M.Sierka}, \citenamefont {R.Wlodarczyk}, \citenamefont {J.Sauer},\ and\
  \citenamefont {Freund}}]{lichtenstein2012}%
  \BibitemOpen
  \bibfield  {author} {\bibinfo {author} {\bibnamefont {L.Lichtenstein}},
  \bibinfo {author} {\bibfnamefont {C.}~\bibnamefont {Buechner}}, \bibinfo
  {author} {\bibnamefont {B.Yang}}, \bibinfo {author} {\bibnamefont
  {S.Shaikhutdinov}}, \bibinfo {author} {\bibnamefont {M.Heyde}}, \bibinfo
  {author} {\bibnamefont {M.Sierka}}, \bibinfo {author} {\bibnamefont
  {R.Wlodarczyk}}, \bibinfo {author} {\bibnamefont {J.Sauer}}, \ and\ \bibinfo
  {author} {\bibfnamefont {H-J}\ \bibnamefont {Freund}},\ }\bibfield  {title}
  {\enquote {\bibinfo {title} {The atomic structure of a metal-supported
  vitreous thin silica film},}\ }\href {\doibase 10.1002/anie.201107097}
  {\bibfield  {journal} {\bibinfo  {journal} {Angewandte Chemie International
  Edition}\ }\textbf {\bibinfo {volume} {51}},\ \bibinfo {pages} {404--407}
  (\bibinfo {year} {2012})}\BibitemShut {NoStop}%
\bibitem [{\citenamefont {P.Y.Huang}\ \emph {et~al.}(2012)\citenamefont
  {P.Y.Huang}, \citenamefont {S.Kurasch}, \citenamefont {A.Srivastava},
  \citenamefont {V.Skakalova}, \citenamefont {J.Kotakoski}, \citenamefont
  {A.V.Krasheninnikov}, \citenamefont {R.Hovden}, \citenamefont {Q.Mao},
  \citenamefont {J.C.Meyer}, \citenamefont {J.Smet}, \citenamefont
  {D.A.Muller},\ and\ \citenamefont {U.Kaiser}}]{huang2012}%
  \BibitemOpen
  \bibfield  {author} {\bibinfo {author} {\bibnamefont {P.Y.Huang}}, \bibinfo
  {author} {\bibnamefont {S.Kurasch}}, \bibinfo {author} {\bibnamefont
  {A.Srivastava}}, \bibinfo {author} {\bibnamefont {V.Skakalova}}, \bibinfo
  {author} {\bibnamefont {J.Kotakoski}}, \bibinfo {author} {\bibnamefont
  {A.V.Krasheninnikov}}, \bibinfo {author} {\bibnamefont {R.Hovden}}, \bibinfo
  {author} {\bibnamefont {Q.Mao}}, \bibinfo {author} {\bibnamefont
  {J.C.Meyer}}, \bibinfo {author} {\bibnamefont {J.Smet}}, \bibinfo {author}
  {\bibnamefont {D.A.Muller}}, \ and\ \bibinfo {author} {\bibnamefont
  {U.Kaiser}},\ }\bibfield  {title} {\enquote {\bibinfo {title} {Direct imaging
  of a two-dimensional silica glass on graphene},}\ }\href {\doibase
  10.1021/nl204423x} {\bibfield  {journal} {\bibinfo  {journal} {Nano letters}\
  }\textbf {\bibinfo {volume} {12}},\ \bibinfo {pages} {1081--1086} (\bibinfo
  {year} {2012})}\BibitemShut {NoStop}%
\bibitem [{\citenamefont {Salmon}\ \emph {et~al.}(2005)\citenamefont {Salmon},
  \citenamefont {Martin}, \citenamefont {Mason},\ and\ \citenamefont
  {Cuello}}]{salmon2005a}%
  \BibitemOpen
  \bibfield  {author} {\bibinfo {author} {\bibfnamefont {Philip~S}\
  \bibnamefont {Salmon}}, \bibinfo {author} {\bibfnamefont {Richard~A}\
  \bibnamefont {Martin}}, \bibinfo {author} {\bibfnamefont {Philip~E}\
  \bibnamefont {Mason}}, \ and\ \bibinfo {author} {\bibfnamefont {Gabriel~J}\
  \bibnamefont {Cuello}},\ }\bibfield  {title} {\enquote {\bibinfo {title}
  {Topological versus chemical ordering in network glasses at intermediate and
  extended length scales},}\ }\href {\doibase 10.1038/nature03475} {\bibfield
  {journal} {\bibinfo  {journal} {Nature}\ }\textbf {\bibinfo {volume} {435}},\
  \bibinfo {pages} {75--78} (\bibinfo {year} {2005})}\BibitemShut {NoStop}%
\bibitem [{\citenamefont {Salmon}(2005)}]{salmon2005b}%
  \BibitemOpen
  \bibfield  {author} {\bibinfo {author} {\bibfnamefont {Philip~S}\
  \bibnamefont {Salmon}},\ }\bibfield  {title} {\enquote {\bibinfo {title}
  {Moments of the {B}hatia--{T}hornton partial pair-distribution functions},}\
  }\href {\doibase 10.1088/0953-8984/17/45/045} {\bibfield  {journal} {\bibinfo
   {journal} {Journal of Physics: Condensed Matter}\ }\textbf {\bibinfo
  {volume} {17}},\ \bibinfo {pages} {S3537} (\bibinfo {year}
  {2005})}\BibitemShut {NoStop}%
\bibitem [{\citenamefont {Salmon}(2007)}]{salmon2007b}%
  \BibitemOpen
  \bibfield  {author} {\bibinfo {author} {\bibfnamefont {Philip~S}\
  \bibnamefont {Salmon}},\ }\bibfield  {title} {\enquote {\bibinfo {title} {The
  structure of tetrahedral network glass forming systems at intermediate and
  extended length scales},}\ }\href {\doibase 10.1088/0953-8984/19/45/455208}
  {\bibfield  {journal} {\bibinfo  {journal} {Journal of Physics: Condensed
  Matter}\ }\textbf {\bibinfo {volume} {19}},\ \bibinfo {pages} {455208}
  (\bibinfo {year} {2007})}\BibitemShut {NoStop}%
\bibitem [{\citenamefont {Salmon}\ \emph {et~al.}(2006)\citenamefont {Salmon},
  \citenamefont {Barnes}, \citenamefont {Martin},\ and\ \citenamefont
  {Cuello}}]{salmon2006}%
  \BibitemOpen
  \bibfield  {author} {\bibinfo {author} {\bibfnamefont {Philip~S}\
  \bibnamefont {Salmon}}, \bibinfo {author} {\bibfnamefont {Adrian~C}\
  \bibnamefont {Barnes}}, \bibinfo {author} {\bibfnamefont {Richard~A}\
  \bibnamefont {Martin}}, \ and\ \bibinfo {author} {\bibfnamefont {Gabriel~J}\
  \bibnamefont {Cuello}},\ }\bibfield  {title} {\enquote {\bibinfo {title}
  {Glass fragility and atomic ordering on the intermediate and extended
  range},}\ }\href {\doibase 10.1103/PhysRevLett.96.235502} {\bibfield
  {journal} {\bibinfo  {journal} {Physical Review Letters}\ }\textbf {\bibinfo
  {volume} {96}},\ \bibinfo {pages} {235502} (\bibinfo {year}
  {2006})}\BibitemShut {NoStop}%
\bibitem [{\citenamefont {Salmon}\ \emph {et~al.}(2007)\citenamefont {Salmon},
  \citenamefont {Barnes}, \citenamefont {Martin},\ and\ \citenamefont
  {Cuello}}]{salmon2007}%
  \BibitemOpen
  \bibfield  {author} {\bibinfo {author} {\bibfnamefont {Philip~S}\
  \bibnamefont {Salmon}}, \bibinfo {author} {\bibfnamefont {Adrian~C}\
  \bibnamefont {Barnes}}, \bibinfo {author} {\bibfnamefont {Richard~A}\
  \bibnamefont {Martin}}, \ and\ \bibinfo {author} {\bibfnamefont {Gabriel~J}\
  \bibnamefont {Cuello}},\ }\bibfield  {title} {\enquote {\bibinfo {title}
  {Structure of glassy {G}e{O}$_2$},}\ }\href {\doibase
  10.1088/0953-8984/19/41/415110} {\bibfield  {journal} {\bibinfo  {journal}
  {Journal of Physics: Condensed Matter}\ }\textbf {\bibinfo {volume} {19}},\
  \bibinfo {pages} {415110} (\bibinfo {year} {2007})}\BibitemShut {NoStop}%
\bibitem [{\citenamefont {Wilson}(2016)}]{wilson2016}%
  \BibitemOpen
  \bibfield  {author} {\bibinfo {author} {\bibfnamefont {Mark}\ \bibnamefont
  {Wilson}},\ }\bibfield  {title} {\enquote {\bibinfo {title} {Structure and
  dynamics in network-forming materials},}\ }\href {\doibase
  10.1088/0953-8984/28/50/503001} {\bibfield  {journal} {\bibinfo  {journal}
  {Journal of Physics: Condensed Matter}\ }\textbf {\bibinfo {volume} {28}},\
  \bibinfo {pages} {503001} (\bibinfo {year} {2016})}\BibitemShut {NoStop}%
\bibitem [{\citenamefont {J.Weissenrieder}\ \emph {et~al.}(2005)\citenamefont
  {J.Weissenrieder}, \citenamefont {S.Kaya}, \citenamefont {J-L.Lu},
  \citenamefont {H-J.Gao}, \citenamefont {S.Shaikhutdinov}, \citenamefont
  {H-J.Freund}, \citenamefont {M.Sierka}, \citenamefont {T.K.Todorova},\ and\
  \citenamefont {J.Sauer}}]{weissenrieder2005}%
  \BibitemOpen
  \bibfield  {author} {\bibinfo {author} {\bibnamefont {J.Weissenrieder}},
  \bibinfo {author} {\bibnamefont {S.Kaya}}, \bibinfo {author} {\bibnamefont
  {J-L.Lu}}, \bibinfo {author} {\bibnamefont {H-J.Gao}}, \bibinfo {author}
  {\bibnamefont {S.Shaikhutdinov}}, \bibinfo {author} {\bibnamefont
  {H-J.Freund}}, \bibinfo {author} {\bibnamefont {M.Sierka}}, \bibinfo {author}
  {\bibnamefont {T.K.Todorova}}, \ and\ \bibinfo {author} {\bibnamefont
  {J.Sauer}},\ }\bibfield  {title} {\enquote {\bibinfo {title} {Atomic
  structure of a thin silica film on a {M}o (112) substrate: a two-dimensional
  network of {SiO}$_4$ tetrahedra},}\ }\href@noop {} {\bibfield  {journal}
  {\bibinfo  {journal} {Physical review letters}\ }\textbf {\bibinfo {volume}
  {95}},\ \bibinfo {pages} {076103} (\bibinfo {year} {2005})}\BibitemShut
  {NoStop}%
\bibitem [{\citenamefont {L{\"o}ffler}\ \emph {et~al.}(2010)\citenamefont
  {L{\"o}ffler}, \citenamefont {Uhlrich}, \citenamefont {Baron}, \citenamefont
  {Yang}, \citenamefont {Yu}, \citenamefont {Lichtenstein}, \citenamefont
  {Heinke}, \citenamefont {B{\"u}chner}, \citenamefont {Heyde}, \citenamefont
  {Shaikhutdinov} \emph {et~al.}}]{loffler2010}%
  \BibitemOpen
  \bibfield  {author} {\bibinfo {author} {\bibfnamefont {Daniel}\ \bibnamefont
  {L{\"o}ffler}}, \bibinfo {author} {\bibfnamefont {John~J}\ \bibnamefont
  {Uhlrich}}, \bibinfo {author} {\bibfnamefont {M}~\bibnamefont {Baron}},
  \bibinfo {author} {\bibfnamefont {Bing}\ \bibnamefont {Yang}}, \bibinfo
  {author} {\bibfnamefont {Xin}\ \bibnamefont {Yu}}, \bibinfo {author}
  {\bibfnamefont {Leonid}\ \bibnamefont {Lichtenstein}}, \bibinfo {author}
  {\bibfnamefont {Lars}\ \bibnamefont {Heinke}}, \bibinfo {author}
  {\bibfnamefont {Christin}\ \bibnamefont {B{\"u}chner}}, \bibinfo {author}
  {\bibfnamefont {Markus}\ \bibnamefont {Heyde}}, \bibinfo {author}
  {\bibfnamefont {Shamil}\ \bibnamefont {Shaikhutdinov}},  \emph {et~al.},\
  }\bibfield  {title} {\enquote {\bibinfo {title} {Growth and structure of
  crystalline silica sheet on {R}u (0001)},}\ }\href {\doibase
  10.1103/PhysRevLett.105.146104} {\bibfield  {journal} {\bibinfo  {journal}
  {Physical review letters}\ }\textbf {\bibinfo {volume} {105}},\ \bibinfo
  {pages} {146104} (\bibinfo {year} {2010})}\BibitemShut {NoStop}%
\bibitem [{\citenamefont {M.Wilson}\ \emph {et~al.}(2013)\citenamefont
  {M.Wilson}, \citenamefont {A.Kumar}, \citenamefont {D.Sherrington},\ and\
  \citenamefont {M.F.Thorpe}}]{wilson2013}%
  \BibitemOpen
  \bibfield  {author} {\bibinfo {author} {\bibnamefont {M.Wilson}}, \bibinfo
  {author} {\bibnamefont {A.Kumar}}, \bibinfo {author} {\bibnamefont
  {D.Sherrington}}, \ and\ \bibinfo {author} {\bibnamefont {M.F.Thorpe}},\
  }\bibfield  {title} {\enquote {\bibinfo {title} {Modeling vitreous silica
  bilayers},}\ }\href {\doibase 10.1103/PhysRevB.87.214108} {\bibfield
  {journal} {\bibinfo  {journal} {Physical Review B}\ }\textbf {\bibinfo
  {volume} {87}},\ \bibinfo {pages} {214108} (\bibinfo {year}
  {2013})}\BibitemShut {NoStop}%
\bibitem [{\citenamefont {Kumar}\ \emph {et~al.}(2014)\citenamefont {Kumar},
  \citenamefont {Sherrington}, \citenamefont {Wilson},\ and\ \citenamefont
  {Thorpe}}]{kumar2014}%
  \BibitemOpen
  \bibfield  {author} {\bibinfo {author} {\bibfnamefont {A.}~\bibnamefont
  {Kumar}}, \bibinfo {author} {\bibfnamefont {D.}~\bibnamefont {Sherrington}},
  \bibinfo {author} {\bibfnamefont {M.}~\bibnamefont {Wilson}}, \ and\ \bibinfo
  {author} {\bibfnamefont {M.~F.}\ \bibnamefont {Thorpe}},\ }\bibfield  {title}
  {\enquote {\bibinfo {title} {Ring statistics of silica bilayers},}\ }\href
  {\doibase 10.1088/0953-8984/26/39/395401} {\bibfield  {journal} {\bibinfo
  {journal} {Journal of Physics: Condensed Matter}\ }\textbf {\bibinfo {volume}
  {26}},\ \bibinfo {pages} {395401} (\bibinfo {year} {2014})}\BibitemShut
  {NoStop}%
\bibitem [{\citenamefont {Lebedev}(1937)}]{lebedev1937}%
  \BibitemOpen
  \bibfield  {author} {\bibinfo {author} {\bibfnamefont {AA}~\bibnamefont
  {Lebedev}},\ }\bibfield  {title} {\enquote {\bibinfo {title} {O polimorfizme
  i otzhige stekla, trud’i gos. opt. inst. 2 1-20 (1921)(in russian);
  ibid},}\ }\href@noop {} {\bibfield  {journal} {\bibinfo  {journal} {Izv.
  Akad. Nauk SSSR, Otd. Mat. Estestv. Nauk, Ser. Fiz}\ }\textbf {\bibinfo
  {volume} {3}},\ \bibinfo {pages} {381} (\bibinfo {year} {1937})}\BibitemShut
  {NoStop}%
\bibitem [{\citenamefont {Keen}\ and\ \citenamefont {Dove}(1999)}]{keen1999}%
  \BibitemOpen
  \bibfield  {author} {\bibinfo {author} {\bibfnamefont {David~A}\ \bibnamefont
  {Keen}}\ and\ \bibinfo {author} {\bibfnamefont {Martin~T}\ \bibnamefont
  {Dove}},\ }\bibfield  {title} {\enquote {\bibinfo {title} {Local structures
  of amorphous and crystalline phases of silica, {S}i{O}$_2$, by neutron total
  scattering},}\ }\href {\doibase 10.1088/0953-8984/11/47/311} {\bibfield
  {journal} {\bibinfo  {journal} {Journal of Physics: Condensed Matter}\
  }\textbf {\bibinfo {volume} {11}},\ \bibinfo {pages} {9263} (\bibinfo {year}
  {1999})}\BibitemShut {NoStop}%
\bibitem [{\citenamefont {Thorpe}(1983)}]{thorpe1983continuous}%
  \BibitemOpen
  \bibfield  {author} {\bibinfo {author} {\bibfnamefont {M~F}\ \bibnamefont
  {Thorpe}},\ }\bibfield  {title} {\enquote {\bibinfo {title} {Continuous
  deformations in random networks},}\ }\href {\doibase
  10.1016/0022-3093(83)90424-6} {\bibfield  {journal} {\bibinfo  {journal}
  {Journal of Non-Crystalline Solids}\ }\textbf {\bibinfo {volume} {57}},\
  \bibinfo {pages} {355--370} (\bibinfo {year} {1983})}\BibitemShut {NoStop}%
\bibitem [{\citenamefont {Ellenbroek}\ \emph {et~al.}(2015)\citenamefont
  {Ellenbroek}, \citenamefont {Hagh}, \citenamefont {Kumar}, \citenamefont
  {Thorpe},\ and\ \citenamefont {van Hecke}}]{ellenbroek2015rigidity}%
  \BibitemOpen
  \bibfield  {author} {\bibinfo {author} {\bibfnamefont {Wouter~G}\
  \bibnamefont {Ellenbroek}}, \bibinfo {author} {\bibfnamefont {Varda~F}\
  \bibnamefont {Hagh}}, \bibinfo {author} {\bibfnamefont {Avishek}\
  \bibnamefont {Kumar}}, \bibinfo {author} {\bibfnamefont {M~F}\ \bibnamefont
  {Thorpe}}, \ and\ \bibinfo {author} {\bibfnamefont {Martin}\ \bibnamefont
  {van Hecke}},\ }\bibfield  {title} {\enquote {\bibinfo {title} {Rigidity loss
  in disordered systems: Three scenarios},}\ }\href {\doibase
  10.1103/PhysRevLett.114.135501} {\bibfield  {journal} {\bibinfo  {journal}
  {Physical review letters}\ }\textbf {\bibinfo {volume} {114}},\ \bibinfo
  {pages} {135501} (\bibinfo {year} {2015})}\BibitemShut {NoStop}%
\bibitem [{\citenamefont {Theran}\ \emph {et~al.}(2015)\citenamefont {Theran},
  \citenamefont {Nixon}, \citenamefont {Ross}, \citenamefont {Sadjadi},
  \citenamefont {Servatius},\ and\ \citenamefont {Thorpe}}]{theran2015}%
  \BibitemOpen
  \bibfield  {author} {\bibinfo {author} {\bibfnamefont {L.}~\bibnamefont
  {Theran}}, \bibinfo {author} {\bibfnamefont {A.}~\bibnamefont {Nixon}},
  \bibinfo {author} {\bibfnamefont {E.}~\bibnamefont {Ross}}, \bibinfo {author}
  {\bibfnamefont {M.}~\bibnamefont {Sadjadi}}, \bibinfo {author} {\bibfnamefont
  {B.}~\bibnamefont {Servatius}}, \ and\ \bibinfo {author} {\bibfnamefont
  {M.F.}\ \bibnamefont {Thorpe}},\ }\bibfield  {title} {\enquote {\bibinfo
  {title} {Anchored boundary conditions for locally isostatic networks},}\
  }\href {\doibase 10.1103/PhysRevE.92.053306} {\bibfield  {journal} {\bibinfo
  {journal} {Physical Review E}\ }\textbf {\bibinfo {volume} {92}},\ \bibinfo
  {pages} {053306} (\bibinfo {year} {2015})}\BibitemShut {NoStop}%
\bibitem [{\citenamefont {Sartbaeva}\ \emph {et~al.}(2007)\citenamefont
  {Sartbaeva}, \citenamefont {Wells}, \citenamefont {Huerta},\ and\
  \citenamefont {Thorpe}}]{sartbaeva2007}%
  \BibitemOpen
  \bibfield  {author} {\bibinfo {author} {\bibfnamefont {A}~\bibnamefont
  {Sartbaeva}}, \bibinfo {author} {\bibfnamefont {SA}~\bibnamefont {Wells}},
  \bibinfo {author} {\bibfnamefont {A}~\bibnamefont {Huerta}}, \ and\ \bibinfo
  {author} {\bibfnamefont {MF}~\bibnamefont {Thorpe}},\ }\bibfield  {title}
  {\enquote {\bibinfo {title} {Local structural variability and the
  intermediate phase window in network glasses},}\ }\href {\doibase
  10.1103/PhysRevB.75.224204} {\bibfield  {journal} {\bibinfo  {journal}
  {Physical Review B}\ }\textbf {\bibinfo {volume} {75}},\ \bibinfo {pages}
  {224204} (\bibinfo {year} {2007})}\BibitemShut {NoStop}%
\bibitem [{\citenamefont {Madden}\ and\ \citenamefont
  {Wilson}(1996)}]{madden1996}%
  \BibitemOpen
  \bibfield  {author} {\bibinfo {author} {\bibfnamefont {Paul~A}\ \bibnamefont
  {Madden}}\ and\ \bibinfo {author} {\bibfnamefont {Mark}\ \bibnamefont
  {Wilson}},\ }\bibfield  {title} {\enquote {\bibinfo {title} {`{C}ovalent'
  effects in `ionic' systems},}\ }\href {\doibase 10.1039/CS9962500339}
  {\bibfield  {journal} {\bibinfo  {journal} {Chemical Society Reviews}\
  }\textbf {\bibinfo {volume} {25}},\ \bibinfo {pages} {339--350} (\bibinfo
  {year} {1996})}\BibitemShut {NoStop}%
\bibitem [{\citenamefont {Tangney}\ and\ \citenamefont
  {Scandolo}(2002)}]{tangney2002}%
  \BibitemOpen
  \bibfield  {author} {\bibinfo {author} {\bibfnamefont {Paul}\ \bibnamefont
  {Tangney}}\ and\ \bibinfo {author} {\bibfnamefont {Sandro}\ \bibnamefont
  {Scandolo}},\ }\bibfield  {title} {\enquote {\bibinfo {title} {An ab initio
  parametrized interatomic force field for silica},}\ }\href {\doibase
  10.1063/1.1513312} {\bibfield  {journal} {\bibinfo  {journal} {The Journal of
  chemical physics}\ }\textbf {\bibinfo {volume} {117}},\ \bibinfo {pages}
  {8898--8904} (\bibinfo {year} {2002})}\BibitemShut {NoStop}%
\bibitem [{\citenamefont {A.Kumar}\ \emph {et~al.}(2012)\citenamefont
  {A.Kumar}, \citenamefont {M.Wilson},\ and\ \citenamefont
  {M.F.Thorpe}}]{kumar2012}%
  \BibitemOpen
  \bibfield  {author} {\bibinfo {author} {\bibnamefont {A.Kumar}}, \bibinfo
  {author} {\bibnamefont {M.Wilson}}, \ and\ \bibinfo {author} {\bibnamefont
  {M.F.Thorpe}},\ }\bibfield  {title} {\enquote {\bibinfo {title} {Amorphous
  graphene: a realization of {Z}achariasen’s glass},}\ }\href {\doibase
  10.1088/0953-8984/24/48/485003} {\bibfield  {journal} {\bibinfo  {journal}
  {Journal of Physics: Condensed Matter}\ }\textbf {\bibinfo {volume} {24}},\
  \bibinfo {pages} {485003} (\bibinfo {year} {2012})}\BibitemShut {NoStop}%
\bibitem [{\citenamefont {Soler}\ \emph {et~al.}(2002)\citenamefont {Soler},
  \citenamefont {Artacho}, \citenamefont {Gale}, \citenamefont {Garc{\'\i}a},
  \citenamefont {Junquera}, \citenamefont {Ordej{\'o}n},\ and\ \citenamefont
  {S{\'a}nchez-Portal}}]{soler2002}%
  \BibitemOpen
  \bibfield  {author} {\bibinfo {author} {\bibfnamefont {Jos{\'e}~M}\
  \bibnamefont {Soler}}, \bibinfo {author} {\bibfnamefont {Emilio}\
  \bibnamefont {Artacho}}, \bibinfo {author} {\bibfnamefont {Julian~D}\
  \bibnamefont {Gale}}, \bibinfo {author} {\bibfnamefont {Alberto}\
  \bibnamefont {Garc{\'\i}a}}, \bibinfo {author} {\bibfnamefont {Javier}\
  \bibnamefont {Junquera}}, \bibinfo {author} {\bibfnamefont {Pablo}\
  \bibnamefont {Ordej{\'o}n}}, \ and\ \bibinfo {author} {\bibfnamefont
  {Daniel}\ \bibnamefont {S{\'a}nchez-Portal}},\ }\bibfield  {title} {\enquote
  {\bibinfo {title} {The siesta method for ab initio order-n materials
  simulation},}\ }\href {\doibase 10.1088/0953-8984/14/11/302} {\bibfield
  {journal} {\bibinfo  {journal} {Journal of Physics: Condensed Matter}\
  }\textbf {\bibinfo {volume} {14}},\ \bibinfo {pages} {2745} (\bibinfo {year}
  {2002})}\BibitemShut {NoStop}%
\bibitem [{\citenamefont {Bhattarai}\ and\ \citenamefont
  {Drabold}(2016)}]{bhattarai2016}%
  \BibitemOpen
  \bibfield  {author} {\bibinfo {author} {\bibfnamefont {Bishal}\ \bibnamefont
  {Bhattarai}}\ and\ \bibinfo {author} {\bibfnamefont {DA}~\bibnamefont
  {Drabold}},\ }\bibfield  {title} {\enquote {\bibinfo {title} {Vibrations in
  amorphous silica},}\ }\href {\doibase 10.1016/j.jnoncrysol.2016.02.002}
  {\bibfield  {journal} {\bibinfo  {journal} {Journal of Non-Crystalline
  Solids}\ }\textbf {\bibinfo {volume} {439}},\ \bibinfo {pages} {6--14}
  (\bibinfo {year} {2016})}\BibitemShut {NoStop}%
\bibitem [{\citenamefont {Desa}\ \emph {et~al.}(1982)\citenamefont {Desa},
  \citenamefont {Wright}, \citenamefont {Wong},\ and\ \citenamefont
  {Sinclair}}]{desa1982}%
  \BibitemOpen
  \bibfield  {author} {\bibinfo {author} {\bibfnamefont {JA~Erwin}\
  \bibnamefont {Desa}}, \bibinfo {author} {\bibfnamefont {Adrian~C}\
  \bibnamefont {Wright}}, \bibinfo {author} {\bibfnamefont {Joe}\ \bibnamefont
  {Wong}}, \ and\ \bibinfo {author} {\bibfnamefont {Roger~N}\ \bibnamefont
  {Sinclair}},\ }\bibfield  {title} {\enquote {\bibinfo {title} {A neutron
  diffraction investigation of the structure of vitreous zinc chloride},}\
  }\href {\doibase 10.1016/0022-3093(82)90189-2} {\bibfield  {journal}
  {\bibinfo  {journal} {Journal of Non-Crystalline Solids}\ }\textbf {\bibinfo
  {volume} {51}},\ \bibinfo {pages} {57--86} (\bibinfo {year}
  {1982})}\BibitemShut {NoStop}%
\bibitem [{\citenamefont {Neuefeind}\ and\ \citenamefont
  {Liss}(1996)}]{neuefeind1996}%
  \BibitemOpen
  \bibfield  {author} {\bibinfo {author} {\bibfnamefont {J{\"o}rg}\
  \bibnamefont {Neuefeind}}\ and\ \bibinfo {author} {\bibfnamefont {K-D}\
  \bibnamefont {Liss}},\ }\bibfield  {title} {\enquote {\bibinfo {title} {Bond
  angle distribution in amorphous germania and silica},}\ }\href {\doibase
  10.1002/bbpc.19961000812} {\bibfield  {journal} {\bibinfo  {journal}
  {Berichte der Bunsengesellschaft f{\"u}r physikalische Chemie}\ }\textbf
  {\bibinfo {volume} {100}},\ \bibinfo {pages} {1341--1349} (\bibinfo {year}
  {1996})}\BibitemShut {NoStop}%
\bibitem [{\citenamefont {Dawson}\ \emph {et~al.}(2014)\citenamefont {Dawson},
  \citenamefont {Sanchez-Smith}, \citenamefont {Rez}, \citenamefont {O'Keefe},\
  and\ \citenamefont {Treacy}}]{dawson2014}%
  \BibitemOpen
  \bibfield  {author} {\bibinfo {author} {\bibfnamefont {C.J.}\ \bibnamefont
  {Dawson}}, \bibinfo {author} {\bibfnamefont {R.}~\bibnamefont
  {Sanchez-Smith}}, \bibinfo {author} {\bibfnamefont {P.}~\bibnamefont {Rez}},
  \bibinfo {author} {\bibfnamefont {M.}~\bibnamefont {O'Keefe}}, \ and\
  \bibinfo {author} {\bibfnamefont {M.M.J.}\ \bibnamefont {Treacy}},\
  }\bibfield  {title} {\enquote {\bibinfo {title} {Ab initio calculations of
  the energy dependence of {Si--O--Si} angles in silica and {Ge--O--Ge} angles
  in germania crystalline systems},}\ }\href {\doibase 10.1021/cm402814v}
  {\bibfield  {journal} {\bibinfo  {journal} {Chemistry of Materials}\ }\textbf
  {\bibinfo {volume} {26}},\ \bibinfo {pages} {1523--1527} (\bibinfo {year}
  {2014})}\BibitemShut {NoStop}%
\bibitem [{\citenamefont {Perrakis}\ \emph {et~al.}(1999)\citenamefont
  {Perrakis}, \citenamefont {Morris},\ and\ \citenamefont
  {Lamzin}}]{perrakis1999}%
  \BibitemOpen
  \bibfield  {author} {\bibinfo {author} {\bibfnamefont {Anastassis}\
  \bibnamefont {Perrakis}}, \bibinfo {author} {\bibfnamefont {Richard}\
  \bibnamefont {Morris}}, \ and\ \bibinfo {author} {\bibfnamefont {Victor~S}\
  \bibnamefont {Lamzin}},\ }\bibfield  {title} {\enquote {\bibinfo {title}
  {Automated protein model building combined with iterative structure
  refinement},}\ }\href {\doibase 10.1038/8263} {\bibfield  {journal} {\bibinfo
   {journal} {Nature structural \& molecular biology}\ }\textbf {\bibinfo
  {volume} {6}},\ \bibinfo {pages} {458--463} (\bibinfo {year}
  {1999})}\BibitemShut {NoStop}%
\bibitem [{\citenamefont {Towns}\ \emph {et~al.}(2014)\citenamefont {Towns},
  \citenamefont {Cockerill}, \citenamefont {Dahan}, \citenamefont {Foster},
  \citenamefont {Gaither}, \citenamefont {Grimshaw}, \citenamefont {Hazlewood},
  \citenamefont {Lathrop}, \citenamefont {Lifka}, \citenamefont {Peterson}
  \emph {et~al.}}]{towns2014}%
  \BibitemOpen
  \bibfield  {author} {\bibinfo {author} {\bibfnamefont {John}\ \bibnamefont
  {Towns}}, \bibinfo {author} {\bibfnamefont {Timothy}\ \bibnamefont
  {Cockerill}}, \bibinfo {author} {\bibfnamefont {Maytal}\ \bibnamefont
  {Dahan}}, \bibinfo {author} {\bibfnamefont {Ian}\ \bibnamefont {Foster}},
  \bibinfo {author} {\bibfnamefont {Kelly}\ \bibnamefont {Gaither}}, \bibinfo
  {author} {\bibfnamefont {Andrew}\ \bibnamefont {Grimshaw}}, \bibinfo {author}
  {\bibfnamefont {Victor}\ \bibnamefont {Hazlewood}}, \bibinfo {author}
  {\bibfnamefont {Scott}\ \bibnamefont {Lathrop}}, \bibinfo {author}
  {\bibfnamefont {Dave}\ \bibnamefont {Lifka}}, \bibinfo {author}
  {\bibfnamefont {Gregory~D}\ \bibnamefont {Peterson}},  \emph {et~al.},\
  }\bibfield  {title} {\enquote {\bibinfo {title} {{XSEDE}: accelerating
  scientific discovery},}\ }\href {\doibase 10.1109/MCSE.2014.80} {\bibfield
  {journal} {\bibinfo  {journal} {Computing in Science \& Engineering}\
  }\textbf {\bibinfo {volume} {16}},\ \bibinfo {pages} {62--74} (\bibinfo
  {year} {2014})}\BibitemShut {NoStop}%
\end{thebibliography}%
\end{document}